\documentstyle[aps,preprint]{revtex}

\newcommand{\Bra}[1]{\left\langle#1\right\vert}
\newcommand{\Ket}[1]{\left\vert#1\right\rangle}
\newcommand{\BraKet}[2]{\left\langle#1\right\vert\left.#2\right\rangle}
\newcommand{\KetBra}[2]{\left\vert#1\right\rangle\left\langle#2\right\vert}
\newcommand{\MatrixEl}[3]{\left\langle#1\right\vert #2 \left\vert#3\right\rangle}
\newcommand{\MeanValue}[1]{\left\langle#1\right\rangle}

\begin{document} \draft \textheight=9in \textwidth=6.5in \draft 

\title{Reconstructing the vibrational state of a trapped ion}
\author{B. Militello\footnote{e-mail: bdmilite@fisica.unipa.it}, A. Napoli, A. Messina\\
        \emph{INFM, MIUR}
        \emph{and Dipartimento di Scienze Fisiche ed Astronomiche} \\
        \emph{via Archirafi 36, 90123 Palermo (ITALY)}
        }
\maketitle

\begin{abstract}
A new approach for reconstructing the vibrational quantum state of a trapped ion is proposed. 
The method rests upon the current ability of manipulating the trapped ion state and on the
possibility of effectively measuring the scalar product of the two vibrational cofactors of a vibronic entangled state. The experimental feasibility of the method is briefly discussed.
\end{abstract}


\pagebreak

\section{Introduction}

In the last few years trapped ions have furnished a very interesting physical scenario wherein fundamental aspects of quantum mechanics may be tested\cite{Vogel-Rass,nist}. In such systems a charged particle, typically an ion, is subjected to a time dependent and inhomogeneous electromagnetic field generating an \emph{effective} tridimensional harmonic potential involving the ion center of mass coordinates. The system we wish to study thus possesses bosonic and fermionic degrees of freedom, the last ones describing electronic motion\cite{Vogel-Rass,nist}.
Acting upon the ion by suitable classical laser fields, it is possible to generate a wide class of interactions involving both bosonic and electronic operators\cite{Vogel-Rass,nist}. This circumstance leads to the possibility of manipulating the vibrational trapped ion state practically at will, generating number, coherent, squeezed and Schr\"odinger cat states\cite{Vogel-Rass,nist}. 

In order to observe nonclassical features of the system, methods for trapped ion quantum state tomography have been proposed and experimentally realized\cite{Davidovich,nist-tomography}. Here we recall the Wigner function reconstruction scheme\cite{Davidovich} and a technique for observing the density operator discrete Fourier transform\cite{nist-tomography}.
In both cases, partial information, for instance a specific Fock state basis quantum coherence or population associated with the ion vibrational state, \emph{is not directly achievable} without reconstructing the quantum state at all.

In this paper we propose a scheme for measuring the Fock state basis matrix elements of the vibrational trapped ion density operator, $\hat{\rho}^{vibr}_{ij}\equiv\MatrixEl{i}{\hat{\rho}^{vibr}}{j}$. Here each matrix element is measurable independently from the measurement of all other coherences and populations. 
This aspect is new and interesting because it allows the recovering of \emph{partial but meaningful information}, in the sense that we can measure single density operator matrix elements getting each time useful information concerning the physical system under scrutiny. 

For instance, just looking at a single coherence might be of relevance to monitor the incoming of decoherence. 
As done in \cite{nist-tomography}, we can indeed compare the time evolution of $|\hat{\rho}^{vibr}_{20}|$ with $\sqrt{\hat{\rho}^{vibr}_{00}\hat{\rho}^{vibr}_{22}}$. 
A virtue of our scheme is that it allows to measure only these three matrix elements avoiding the complete reconstruction of the density operator.
Of course, measuring \emph{all} density operator matrix elements furnishes the complete reconstruction of the system state. 
From a practical point of view, only few phonons are generally involved in the dynamics of trapped ions, so that what we really need for complete state reconstruction is the individuation of the matrix elements placed just in a square matrix of finite dimensions. 

In the following we analyse a new tomographic technique in the case of a pure state. Such a method is based upon the idea that measuring transverse Pauli operators, $\hat{\sigma}^{\pm}_{x}$ and $\hat{\sigma}^{\pm}_{y}$, when the system is in the entangled state $\frac{1}{\sqrt{2}}\left[\Ket{\phi_{+}}\Ket{+}+\Ket{\phi_{-}}\Ket{-}\right]$, gives as result the real and imaginary parts of the overlap $\BraKet{\phi_{+}}{\phi_{-}}$.
Subsequently we will briefly show that the same protocol remains valid also in the case of non pure states for measuring quantum coherences.

\section{Measuring coherences in trapped ions}

The physical system on which we focus is a three-level trapped ion into a two-dimensional trap, described by the free Hamiltonian
\begin{equation} \label{Unperturbed}
  \hat{H}_0=\sum_{k=x,z}\hbar\omega_{k}\hat{a}_k^{\dag}\hat{a}_k
           +\sum_{l=\pm,\xi}\hbar\omega_{l}\KetBra{l}{l}
\end{equation}
and prepared in the state
\begin{equation} \label{InitialState}
  \Ket{\psi}=\Ket{\phi}_x\Ket{0}_z\Ket{-}
\end{equation}
Here we introduce the Pauli operators related to each two-dimensional subspace of the total tridimensional electronic space $\left\{ \Ket{-}, \Ket{+}, \Ket{\xi} \right\}$: $\hat{\sigma}^{lj}_{x}=\KetBra{l}{j}+\KetBra{j}{l}$, $\hat{\sigma}^{lj}_{y}=i(\KetBra{l}{j}-\KetBra{j}{l})$ and $\hat{\sigma}^{lj}_{z}=\KetBra{j}{j}-\KetBra{l}{l}$, with $l,j=\pm,\xi$.

Our target is to measure quantum coherences related to the vibrational state $\Ket{\phi}$ describing the $x$ mode in eq.(\ref{InitialState}): $\BraKet{\phi}{n}\BraKet{m}{\phi}$.
To this end, assume we are able to generate a unitary transformation $\hat{U}_{mn}$ such that
\begin{equation} \label{Transformation}
  \Ket{{\psi^{mn}}}=
    \hat{U}_{mn}\Ket{\psi}=
    \frac{1}{\sqrt{2}}\left[\Ket{\phi}_x\Ket{m}_z\Ket{-}+\Ket{n}_x\Ket{\phi}_z\Ket{+}\right]
\end{equation}

It is easy to prove that, in this state, expectation values of the transverse pseudospin operators, 
$\hat{\sigma}^{\pm}_{x}\equiv\hat{\sigma}^{-+}_{x}$ and 
$\hat{\sigma}^{\pm}_{y}\equiv\hat{\sigma}^{-+}_{y}$, 
give our target quantity
\begin{equation}\label{TwoModeOverlap}
  \MeanValue{\hat{\sigma}^{\pm}_{x}}+i\MeanValue{\hat{\sigma}^{\pm}_{y}}=\BraKet{\phi}{n}\BraKet{m}{\phi}
\end{equation}
which, in the case of pure states, coincides with a $mn$ Fock state basis coherence. Thus we need to perform $\hat{U}_{mn}$.
Such a unitary transformation may be thought of as the product of three canonical transformations, 
\begin{equation}\label{UnitaryOp}
  \hat{U}_{mn}=\hat{V}^{+}_{n}\hat{V}^{-}_{m}\hat{U}_{00}
\end{equation}
where $\hat{U}_{00}$ is a particular case of $\hat{U}_{mn}$ with $m=n=0$, while $\hat{V}^{+}_{k}$ ($\hat{V}^{-}_{k}$) transforms the state 
$\Ket{0}_x\Ket{\phi}_z\Ket{+}$ ($\Ket{\phi}_x\Ket{0}_z\Ket{-}$) into 
$\Ket{k}_x\Ket{\phi}_z\Ket{+}$ ($\Ket{\phi}_x\Ket{k}_z\Ket{-}$). 
The transformation $\hat{V}^{+}_{k}$ ($\hat{V}^{-}_{k}$) may be easily experimentally realized following standard techniques of generation of Fock states\cite{nist} just involving as electronic levels $\Ket{+}$ ($\Ket{-}$) and $\Ket{\xi}$, for instance via the Hamiltonian models $\hat{H}^{+\xi}_{jc}=\hbar\gamma_{jc}\left[\hat{a}^{\dag}_x\KetBra{+}{\xi}+\hat{a}_x\KetBra{\xi}{+}\right]$, $\hat{H}^{+\xi}_{ajc}=\hbar\gamma_{ajc}\left[\hat{a}_x\KetBra{+}{\xi}+\hat{a}^{\dag}_x\KetBra{\xi}{+}\right]$ and 
$\hat{H}^{+\xi}_{carr}=\hbar\gamma_{carr}\left[\KetBra{+}{\xi}+\KetBra{\xi}{+}\right]$.

The operator $\hat{U}_{00}$ may also be cast in the form of product of four unitary transformations, as follows
\begin{equation} \label{TransformationForm}
  \hat{U}_{00}=R^{-}\left(\frac{\pi}{4}\right)
               R^{vibr}\left(\frac{\pi}{2}\right)
               R^{+}\left(-\frac{\pi}{4}\right)
               R^{-}\left(\frac{\pi}{4}\right)
\end{equation}
where $R^{l}\left(\theta\right)=e^{i\theta\hat{\sigma}^{l\xi}_{y}}$ with $l=\pm$, while $R^{vibr}\left(\theta\right)=e^{i\theta\hat{L}_y\hat{\sigma}^{+\xi}_{x}}$.
Time evolutions described by $R^{l}\left(\theta\right)$ are non dispersive coherent population inversions commonly realized via Raman pulses tuned to the atomic transition frequencies. The interaction Hamiltonian $\hat{H}_{rot}=\hbar\gamma\hat{L}_y\hat{\sigma}^{+\xi}_{x}$, generating the transformation $R^{vibr}\left(\theta\right)$, may be implemented for example following the scheme presented in ref\cite{Knight}. All these interactions are conceived in the Lamb-Dicke regime, meaning that the wavelengths of the Raman lasers are assumed much larger than the dimensions of the region wherein the ion motion is confined.

The effect of $R^{+}\left(-\frac{\pi}{4}\right)R^{-}\left(\frac{\pi}{4}\right)$ 
on the initial state given by eq.(\ref{InitialState}) is the following transformation: $\Ket{\psi}\rightarrow\Ket{\phi}_x\Ket{0}_z\frac{1}{\sqrt{2}}\left[\Ket{-}+\Ket{\alpha}\right]$ where $\hat{\sigma}^{+\xi}_{x}\Ket{\alpha}=\Ket{\alpha}$. 
Then, acting by $R^{vibr}\left(\frac{\pi}{2}\right)=e^{\hbar\gamma\hat{L}_y\hat{\sigma}^{+\xi}_{x}}$, the state $\Ket{\phi}_x\Ket{0}_z\Ket{-}$ is left unchanged, while the state $\Ket{\phi}_x\Ket{0}_z\Ket{\alpha}$ is \emph{vibrationally rotated} about $y$ of $\frac{\pi}{2}$. In this way we reach $\frac{1}{\sqrt{2}}\left[\Ket{\phi}_x\Ket{0}_z\Ket{-}+\Ket{0}_x\Ket{\phi}_z\Ket{\alpha}\right]$. Finally, the operator $R^{-}\left(\frac{\pi}{4}\right)$ restores $\Ket{+}$ from $\Ket{\alpha}$, giving $\Ket{\psi^{00}}$ defined in eq.(\ref{Transformation}).

Until now we have shown the way to implement the unitary transformation in eq.(\ref{Transformation}) such that the relation in eq.(\ref{TwoModeOverlap}) holds. Moreover, we dealt with just pure states. Nevertheless, it is straightforward to prove that the result presented holds also for non pure states. This means that, if the initial state is
\begin{equation}
  \hat{\rho}=\hat{\rho}^{vibr}\Ket{0}_{zz}\Bra{0}\KetBra{-}{-}
\end{equation}
we can canonically transform it reaching the density operator
\begin{equation}
  \hat{\rho}^{mn}=\hat{U}_{mn}\hat{\rho}_0\hat{U}^{\dag}_{mn}
\end{equation}
which satisfies the following relation,
\begin{equation}\label{NonPureCoherence}
  Tr\{\hat{\rho}^{mn}\hat{\sigma}^{\pm}_{x}\}+i\;Tr\{\hat{\rho}^{mn}\hat{\sigma}^{\pm}_{y}\}=\hat{\rho}^{vibr}_{mn}
\end{equation}

This is the most general statement of the result expressed by eq.(\ref{TwoModeOverlap}) in the particular case of pure states.

\section{Conclusions}

Summarizing we have presented a scheme for measuring quantum coherences related to the state of one vibrational mode of a trapped ion center of mass. To this end we need two vibrational modes (the target and an auxiliary one) and three atomic effective levels.
Two levels ($\Ket{\pm}$) are necessary to performs the scalar product as above described, and a third one ($\Ket{\xi}$) is used for quantum state manipulation to pass from the initial configuration to a suitable quantum state. 

The protocol, apart from possible limitations stemming from the weakness of the coupling constant in the two mode Lamb-Dicke regime Hamiltonian model $\hat{H}_{rot}=\hbar\gamma\hat{L}_y\hat{\sigma}^{+\xi}_{x}$, has however a good degree of experimental feasibility. This is due to the fact that both \lq\lq three level ions\rq\rq and \lq\lq two bosonic (vibrational) modes\rq\rq are conditions in the grasp of experimentalists\cite{ThankToWineland}.

We emphasise that our proposal has the new and interesting feature that it allows the reconstruction of the density operator \lq\lq step by step\rq\rq, doing a set of \emph{meaningful} steps, in the sense that we measure a density operator matrix element a time, taking in mind that each one of these quantities is itself an information concerning the system.

 
\section*{Acknowledgement}
 
We thank S.Maniscalco for stimulating discussions and D.Wineland for useful information about the experimental feasibility of the method.
One of the authors (A. N.) acknowledges financial support from Finanziamento Progetto Giovani Ricercatori anno 1999, Comitato 02.


\begin{thebibliography}{99}  
\bibitem{Vogel-Rass} W. Vogel and S. Wallentowitz, {\it Manipulation of the quantum state of a trapped ion, in Coherence and statistics of photons and atoms}, edited by Jan Perina (Wiley, New York, 2001)
  \bibitem{nist}  D.J. Wineland et al., {\it J. Res. Natl. Inst. Stand. Technol.} {\bf 103} (1998) 259 
  \bibitem{Davidovich} L.G.Lutterbach and L.Davidovich, {\it Phys. Rev. Lett.} {\bf 78} (1997) 2547 
  \bibitem{nist-tomography} D. Leibfried et al. {\it Phys. Rev. Lett.} {\bf 77} (1996)) 4281  
  \bibitem{Knight} J.Steinbach et al., {\it Phys. Rev. A} {\bf 56} (1997) 4815 
  \bibitem{ThankToWineland} D.J. Wineland, \emph{private communications}
\end{thebibliography}
\end{document}